\documentclass[aps,prd,preprint,showkeys]{revtex4}
\usepackage{graphicx}
\usepackage{amsfonts,amsmath,amssymb,bm}
\def\be{\begin{equation}}
\def\te{\end{equation}}
\def\bea{\begin{eqnarray}}
\def\tea{\end{eqnarray}}
\def\nn{\nonumber\\}
\usepackage{color}
\usepackage[sort&compress]{natbib}

\graphicspath{{../figuras_finales/}}

\begin{document}

\title{Magnetothermal instabilities in magnetized anisotropic plasmas}

\author{M. S. Nakwacki}
\email{sole@astro.iag.usp.br}
\affiliation{Instituto de Astronom\'ia, Geof\'isica e Ci\^encias Atmosf\'ericas, Universidade de S\~ao Paulo, 
Rua do Mat\~ao 1226, Cidade Universit\'aria, 05508-090 S\~ao Paulo, Brazil}

\author{J. Peralta-Ramos}
\email{jperalta@ift.unesp.br}

\affiliation{Instituto de F\'isica Te\'orica, Universidade Estadual Paulista, 
Rua Doutor Bento Teobaldo Ferraz 271 - Bloco II, 01140-070 S\~ao Paulo, Brazil}

\date{\today}

\begin{abstract}
Using the transport equations for an ideal anisotropic collisionless plasma derived from the Vlasov equation by the 16--moment method, we analyse the influence of pressure anisotropy exhibited by collisionless magnetized plasmas on the magnetothermal (MTI) and heat--flux--driven buoyancy (HBI) instabilities. 
We calculate the dispersion relation and the growth rates for these instabilities in the presence of a background heat flux and for configurations with static pressure anisotropy, finding that when the frequency at which heat conduction acts is much larger than any other frequency in the system (i.e. weak magnetic field) the pressure anisotropy has no effect on the MTI/HBI, provided the degree of anisotropy is small. In contrast, when this ordering of timescales does not apply the instability criteria depend on pressure anisotropy. Specifically, the growth time of the instabilities in the anisotropic case can be almost one order of magnitude smaller than its isotropic counterpart. 
We conclude that in plasmas where pressure anisotropy is present the MTI/HBI are modified. However, in environments with low magnetic fields and small anisotropy such as the ICM the results obtained from the 16--moment equations under the approximations considered are similar to those obtained from ideal MHD.

\end{abstract}


\maketitle

\section{Introduction}

In many magnetized dilute astrophysical plasmas thermal electron conduction occurs
almost exclusively parallel to magnetic field lines. In this regime,
the equations of ideal magnetohydrodynamics (MHD) that describe the plasma dynamics must
be supplemented with anisotropic transport terms for energy and momentum due to the near
free-streaming motions of particles along magnetic field lines \citep{Braginskii65}. 
\citet{balbus00,balbus04} have shown that anisotropic thermal conduction can fundamentally 
alter the Schwarzschild convective instability criterion in stratified atmospheres, 
in that convection sets in when a temperature gradient (as opposed to an entropy gradient) is present. 
This new convective instability is termed the magnetothermal instability (MTI) \citep{balbus00} when the temperature
decreases with height ($\vec{g}\cdot \vec{\nabla} T > 0$). Later on, \citet{quato} showed that in the presence of a background heat flux there appears a heat--flux--driven buoyancy instability (HBI) when the temperature increases with height ($\vec{g}\cdot \vec{\nabla} T < 0$). Essentially, the MTI and the HBI occur because the heat flux must follow the perturbed magnetic field lines, and thus there are regions in the plasma which are locally heated or cooled  \citep[see][]{balbus00,balbus04,balbus10,quato,mccourt10,parrish06}. 

The MTI and HBI couple the magnetic 
structure of the plasma to its thermal properties and can have important implications 
for galaxy clusters \citep{parrish07,parrish06,parrish08,parrish09,bogdanovic09,sharma09,parrish10,bogdanovic10,ruszkowski10}.
In a weakly magnetized non--rotating atmosphere local simulations have demonstrated that the MTI can amplify
magnetic field and lead to a substantial heat flux down the temperature gradient \citep{parrish05,parrish07}. 
\citet{mccourt10,parrish06,parrish08q,parrish07} studied the non--linear 
development of the MTI and HBI using numerical simulations applied to clusters of galaxies. \citet{mccourt10} found that the HBI can reduce the conductive heat flux through the plasma due to the reorientation of the magnetic field lines. As the HBI should operate
in the innermost $\sim$100--200 kpc in the intracluster medium (ICM) of cool-core galaxy clusters, where the observed temperature
increases outward, the cooling time of the ICM is shorter than its age; so the HBI removes thermal conduction as a source of energy for the cores, increasing the cooling flow problem \citep{parrish09}. They also studied turbulence in the ICM and suggested that the interaction between
turbulence and the HBI might be part of a feedback loop for
the thermal evolution of the ICM \citep[see also][]{parrish10,ruszkowski10}. These authors found that unlike the HBI, the MTI drives strong turbulence
and operates as an effcient magnetic dynamo, much more
akin to adiabatic convection, and that while the MTI cannot saturate by reorienting the magnetic field, it can saturate by making the plasma isothermal. \citet{sharma10} studied thermal instabilities including adiabatic cosmic rays. \citet{balbus10} showed that magnetic field configurations that nominally stabilize the HBI or the MTI can lead to g--mode overstabilities. 

Spherical accretion flow is also subject to the MTI. In this sense \citet{sharma08} investigated the effects of MTI on spherical accretion flow using
global simulations \citep[see also][]{bu10}. Consistent with previous local simulations, they found amplification of the magnetic
field and alignment of field lines with the radial direction (temperature gradient direction).
Other scenarios such as the interiors and surface layers of white dwarfs and neutron stars have been also investigated \citep[see e.g.][]{chang10}. 


Dilute magnetized plasmas not only exhibit anisotropic thermal conduction but also an anisotropic pressure tensor, which results from different kinetic temperatures for electron motion in the parallel and perpendicular direction to the magnetic field. For collisionless plasmas the pressure does not isotropize, and hence if we insist on using a fluid description the standard magnetohydrodynamics theory (MHD) must be modified \citep{cgl,krall,oraevski,barakat82,sharmaPhD,boyd,howes06,ramos,ramos05,ramos07}. 

The purpose of this work is to study the MTI and the HBI in a dilute magnetized plasma 
including the effect of pressure anisotropy. Our main goal is to determine the impact of these anisotropies on the instability criterion for 
the MTI and the HBI. To this end, we employ the theory of kinetic MHD (KMHD) as derived by the 16--moment method from the Vlasov equation \citep{oraevski,ramos,barakat82}. 
This formalism extends the well--known double--adiabatic theory of Chew, Goldenberg and Low (CGL) \citep{cgl} in that it allows for nonvanishing (parallel and perpendicular) heat fluxes, which are neglected in the CGL theory. Moreover, the growth rates and instability criteria for hydromagnetic wave 
propagation obtained from the KMHD equations are in better agreement with those obtained from kinetic theory \citep[see for instance,][]{dza1,dza2,andre}. 

Although we keep the discussion quite general throughout the paper, we have in mind an astrophysical magnetized collisionless plasma, and as a typical example we consider the 
ICM (see section \ref{icm}). In particular, in the ICM one of the main open questions is the amplification of the primordial magnetic field. This environment results unstable 
to the MTI on scales of ten kiloparsecs and larger outside cooling cores (regions where the temperature decreases outward). \citet{parrish08} have shown that 
the MTI can produce convective motions and a magnetic dynamo, leading to an efficient transport of heat. The analysis and results presented here may also be of relevance in similar astrophysical environments.

The paper is organized as follows. In section \ref{basic} we give a brief overview of the formalism we use to describe anisotropic plasmas. In section \ref{mti} we describe the equilibrium state of the anisotropic plasma together with the linearized system of equations. We then present and discuss our results in sections \ref{boussi} and \ref{icm}, and finally we give a brief summary in section \ref{con}. 

\section{Basic equations}
\label{basic}
In this section we briefly review the formalism of KMHD and set the stage for the analysis of wave instabilities given later on. More complete accounts of KMHD and similar approaches (as well as their relation to the kinetic theory of dilute plasmas) can be found in \citep{barakat82,sharmaPhD,boyd,ramos,ramos05,ramos07,howes06}, while an application of the 16--moment equations to the study of the firehose and mirror instabilities can be found in \citep{dza1,dza2}. 

In a magnetized collisionless plasma such as those encountered in the interplanetary and intracluster mediums the pressure tensor is anisotropic with respect to the direction of the magnetic field. Thus, the transverse and longitudinal kinetic particle temperatures (associated with motion in the direction perpendicular and parallel to the magnetic field) will differ from each other, $T_\parallel \neq T_\perp $. Due to the complexity of the kinetic equation for such a plasma, it is often convenient to employ a fluid description, which will naturally differ from the standard MHD description. In general, the single--fluid equations derived by Chew, Golderberg and Low (CGL) \citep{cgl} are used. The energy equation of isotropic MHD is replaced by the double--adiabatic laws (or by the double--polytropic laws in the phenomenological approach of \citet{hau93}). Many studies of wave instability problems are based on the CGL equations \citep[see e.g.][]{hasegawa,hau07}. However, when compared to the results obtained from kinetic theory there is a discrepancy in the criterion for the slow-mode mirror instability, as well as basic differences between the nonlinear evolution of the mirror and firehose instabilities \citep{barbara,dza1,dza2,andre}. 

The inadequacies of the CGL approach stem from the unwarranted neglect of the third--order moment of Vlasov equation. The 16--moment method, which is a generalization of Grad's 13--moment method, provides a consistent way of deriving from the Vlasov equation the correct fluid equations for a heat--conducting anisotropic magnetized plasma. If the Larmor radius is much smaller than the other characteristic lengths of the plasma the equations can be considerably simplified \citep{oraevski} \citep[see also,][]{barakat82,ramos,ramos05,ramos07}. The equations describing the collisionless magnetized plasma then read 
\bea
\label{geneqs1}
\frac{d\rho}{dt} + \rho \vec{\nabla}\cdot \vec{v} = 0 \\
\rho \frac{d \vec{v}}{dt} + \vec{\nabla} \bigg( p_\perp + \frac{B^2}{8\pi}\bigg) 
- \frac{1}{4\pi} (\vec{B}\cdot \vec{\nabla})\vec{B} = \rho \vec{g} + \hat{b}\nabla_\parallel\Delta p  \\
+ \Delta p \bigg[ \hat{b} (\vec{\nabla}\cdot \hat{b}) + \nabla_\parallel\hat{b}\bigg]  \nn
\frac{d\vec{B}}{dt} + \vec{B} (\vec{\nabla}\cdot \vec{v}) - (\vec{B}\cdot \vec{\nabla})\vec{v} = 0 \\ 
\rho T_\parallel\frac{d}{dt}s_\parallel =  -\vec{\nabla} \cdot \vec{Q}_\parallel \\ 
\rho T_\perp\frac{d}{dt}s_\perp =  -\vec{\nabla} \cdot \vec{Q}_\perp 
\label{geneqs}
\tea
where 
\be 
\frac{d}{dt} = \frac{\partial }{\partial t} + (\vec{v}\cdot \vec{\nabla}) ~~ ,
\te 
$\Delta p = p_\perp - p_\parallel$, $\hat{b}=\vec{B}/B$, $\nabla_\parallel = \hat{b}\cdot \vec{\nabla}$ and $\vec{Q}_{\parallel,\perp}$ are the parallel and perpendicular heat fluxes, which we assume for simplicity to be given by Braginskii's approximation \citep{Braginskii65}
\be
\vec{Q}_{\parallel,\perp} = - \hat{b} \chi \nabla_\parallel T_{\parallel,\perp}  
\te 
where $\chi$ is the electron thermal diffusivity. Although $\chi$ varies with temperature as 
\be 
\chi = 6 \times 10^{-7} T^{5/2} \textrm{ergs}~ \textrm{cm}^{-1}~ \textrm{K}^{-1}
\te
for simplicity and following \citet{quato} we will consider it as a constant in what follows.

The specific entropies associated with parallel and perpendicular motion are given by \citep{barbara} 
\be
s_\parallel = \frac{c_V}{3}\ln \bigg(\frac{p_\parallel B^2}{\rho^3}\bigg) ~~ \textrm{and} ~~ 
s_\perp = \frac{2c_V}{3}\ln \bigg(\frac{p_\perp}{\rho B}\bigg)
\te
where $c_V$ is the specific heat. Note that the total entropy 
\be 
s=s_\parallel + s_\perp = c_V \ln \bigg(\frac{p_\parallel^{1/3}p_\perp^{2/3}}{\rho^{5/3}}\bigg)
\te 
reduces to the ordinary MHD expression for the specific entropy in the isotropic case $p_\parallel = p_\perp$. 
It will prove convenient for the problem at hand to employ the entropy production equations for $s_\parallel\pm s_\perp$, instead of those for $s_\parallel$ and $s_\perp$ separately. 

\section{MTI/HBI in the presence of pressure anisotropy}
\label{mti}

In the following we ignore the ion contribution to the conductive heat
flux, which is smaller than the electron contribution by a 
factor of $\sqrt{m_i/m_e} \approx $ 42. As stated in the previous section,  
we assume that the electrons have mean free paths much longer than their Larmor radius 
(as occurs, for example, in the ICM). Under these conditions  
the thermal conductivity of the plasma is strongly anisotropic. 

We consider a thermally stratified plasma in the presence of gravity $\vec{g}=-g\hat{z}$, so that in equilibrium 
\be 
\frac{dp}{dz}  = -\rho g ~.
\te 
The magnetic field of the equilibrium state is assumed to be homogeneous and in the $(x,z)$ plane 
\be 
\vec{B} = B_x \hat{x} + B_z \hat{z} 
\te
so that there is a background heat flux given by
\be 
\vec{Q}_{\parallel,\perp} = -\chi (b_x b_z \hat{x}+ b_z^2 \hat{z})\frac{dT_{\parallel,\perp}}{dz} ~.
\te 
As in \citet{quato}, we assume a steady--state initial equilibrium so that $\vec{\nabla}\cdot \vec{Q} = 0$, which implies a linear temperature profile in $\hat{z}$. Note that $p_\perp = p_\parallel + \textrm{const.}$ and that we consider a vanishing initial velocity. This equilibrium state is a solution to eqs. (\ref{geneqs1})--(\ref{geneqs}).  

Under these circumstances, putting $\delta v_z \sim \textrm{exp}(-i\omega t + i \vec{k}\cdot \vec{r})$ and similarly for other quantities the linearly perturbed equations can be written as
\bea
\label{genlin1}
 \omega \delta \rho - \rho k_v = 0 &&\\
 \omega [k_v\vec{k} - k^2 \delta \vec{v}] - (v_A^2-\Delta p/\rho)k^2 \tilde{b}  \frac{\delta \vec{B}}{B}  &&\nn
 + [\vec{k}\tilde{b} - k^2 \hat{b}]\tilde{b}(\delta p_\perp - \delta p_\parallel) + i (k_z \vec{k} - k^2 \hat{z})g \delta \rho/\rho = 0 &&\\ 
 \omega \delta \vec{B} - \vec{B}k_v + (\vec{k}\cdot \vec{B}) \delta \vec{v} = 0 &&\\
 i\frac{c_V}{3}  \omega\bigg[\frac{1}{p_\parallel}(\delta p_{\parallel}+2\delta p_{\perp})-\frac{\delta\vec{B}}{B}(1-\alpha)+\frac{\delta\rho}{\rho}(3+2\alpha)\bigg]&& \nn 
+\bigg(\frac{ds_\parallel}{dz}+\alpha \frac{ds_\perp}{dz}\bigg)\delta v_z 
-i \chi \tilde{b} \frac{D}{\rho}(\frac{d\ln T_\parallel}{dz} +\alpha \frac{d\ln T_\perp}{dz}) &&\nn 
+ \chi \frac{\tilde{b}^2}{\rho} \bigg(\frac{\delta T_\parallel}{T_\parallel} +\alpha \frac{\delta T_\perp}{T_\perp}\bigg) = 0 &&\\
 i\frac{c_V}{3}  \omega\bigg[\frac{1}{p_\parallel}(\delta p_{\parallel}-2\delta p_{\perp})-\frac{\delta\vec{B}}{B}(1+\alpha)+\frac{\delta\rho}{\rho}(3-2\alpha)\bigg] &&\nn +\bigg(\frac{ds_\parallel}{dz}-\alpha \frac{ds_\perp}{dz}\bigg)\delta v_z
-i \chi \tilde{b} \frac{D}{\rho}(\frac{d\ln T_\parallel}{dz} -\alpha \frac{d\ln T_\perp}{dz}) &&\nn 
+ \chi \frac{\tilde{b}^2}{\rho} \bigg(\frac{\delta T_\parallel}{T_\parallel} -\alpha \frac{\delta T_\perp}{T_\perp}\bigg) = 0
\label{genlin}
\tea
where we have used that 
\be 
\delta \vec{Q}_{\parallel,\perp}= -\chi \delta \hat{b} \nabla_\parallel T_{\parallel,\perp} 
-\chi \hat{b} (\delta \hat{b}\cdot \nabla T_{\parallel,\perp}) - i \chi \hat{b}\tilde{b}\delta T_{\parallel,\perp} ~,
\te 
$v_A = B/\sqrt{4\pi \rho}$ is the Alfv\'en speed, and we have defined 
\bea 
\tilde{b} &=& \vec{k}\cdot \hat{b}\\
k_v &=& \vec{k}\cdot \delta \vec{v} \\
D &=& \bigg[ \frac{\delta B_z}{B} - 2 \frac{\delta B}{B} \bigg] \\
\alpha &=& \frac{T_\perp}{T_\parallel} ~.
\tea 

\section{Neglecting pressure perturbations}
\label{boussi}

We will now consider the case in which pressure perturbations are neglected (Boussineq approximation). The growing modes of interest have growth times much longer than the sound crossing time of the perturbation, so it is sufficient to work in the Boussinesq approximation. In this subsection we closely follow \citet{balbus00} and \citet{quato}. 

From eqs. (\ref{genlin1})--(\ref{genlin}) in the Boussineq approximation we get the following dispersion relation
\bea
\label{dispB}
i A_3 \omega^3 + A_2 \omega^2 + (i A_1+ \tilde{A}_1) \omega + A_0 = 0
\tea
where
\bea 
A_0 &=& (\omega_A^2 - \omega_s^2)A_2 \\
&-& g K  p_\parallel \omega_{c,\parallel}(1+\alpha)\frac{d\ln T}{dz}\nonumber \\
A_1 &=&  \frac{2}{5}\rho c_V T_\parallel (1+\frac{2}{3}\alpha)(\omega_A^2 - F) - \rho N^2 \frac{k_\perp^2}{k^2}\\
\tilde{A}_1 &=& \frac{4}{15}g T_\parallel c_V (1-\alpha) \tilde{b} \frac{k_\perp}{k^2} (b_x k_z - b_z k_x) \\
A_2 &=& p_\parallel (1+\alpha) \omega_{c,\parallel}  \\
A_3 &=& \frac{2}{5}\rho c_V T_\parallel \bigg(1+\frac{2}{3}\alpha \bigg)
\tea 
and 
\bea
\omega_A &=& \vec{k}\cdot \vec{v_A} \\
N^2 &=& \frac{2}{5} g \frac{d}{dz}(s_\parallel+\alpha s_\perp) \\
\omega_{c,\parallel} &=& \frac{2}{5}\chi \frac{T_{\parallel}}{p_\parallel} \tilde{b}^2 
\tea
are the Alfv\'en, Brunt--V\"ais\"al\"a and conduction frequencies, respectively, suitably modified by the anisotropy parameter, and 
$F = (c_{s,\parallel}^2-c_{s,\perp}^2) \tilde{b}^2 $ with $c_{s,\parallel,\perp}^2 = p_{\parallel,\perp}/\rho$.  In eq. (\ref{dispB}) we have put 
\be
K = \frac{1}{k^2}[(1-2b_z^2) k_{\perp}^2+2b_zb_xk_z k_x]
\te
for notational simplicity. To derive eq. (\ref{dispB}) we have used that $\delta \rho/\rho = -\delta T_\parallel /T_\parallel = -\delta T_\perp/T_\perp$ \citep[see][]{quato,balbus00}, and also that $d\ln T_\parallel/dz = d\ln T_\perp/dz = d\ln T/dz$, where $T=(T_\parallel + 2T_\perp)/3$ is the total temperature, which follows since $\alpha$ is assumed to be a constant. 

It is interesting to analyse the special case of a weak magnetic field, in which there exists an ordering in frequencies given by \citep{quato}
\be
\omega_c \gg \omega_d = \bigg(\frac{g}{H}\bigg)^{1/2} \gg \omega_A 
\label{ord}
\te 
where $H$ is the local scale-height of the system and $\omega_d$ its dynamical frequency. Besides, if $\alpha$ is sufficiently close to $1$ then one can safely put $|F| < \omega_c$. In this limit the dispersion relation becomes
\be
\omega^2 \simeq gK\frac{d\ln T_\parallel}{dz} = gK\frac{d\ln T}{dz}
\te
This expression is identical to the result of \citet{quato}, which shows that when the magnetic field is sufficiently weak so that the timescale ordering given in eq. (\ref{ord}) holds and  $\alpha\sim 1$ (as occurs e.g. in the ICM), the MTI and the HBI become independent of the pressure anisotropy. This is one of the main results of this work. 

Following \citet{balbus01,balbus10}, we will now analyse the stability of solutions to eq. (\ref{dispB}) using the Routh--Hurwitz criteria \citep{gantmacher}. It should be noted that the polynomial (\ref{dispB}) has complex coefficients  and therefore the generalized Routh--Hurwitz theorem must be used \footnote{Performing the transformation $\omega \rightarrow i\sigma$ on (\ref{dispB}) results in a polynomial with real coefficients $(A_0,A_1,A_2,A_3)$ but the term containing $\tilde{A}_1$ becomes imaginary.}. Briefly stated, the procedure to determine if a given complex polynomial $f(z)$ with no roots in the imaginary axis is stable (i.e. all of its roots have negative real parts) is as follows \citep[for a detailed account see e.g.][]{gantmacher}: (i) determine the real polynomials $P(r)$ and $Q(r)$ such that $f(ir) = P(r) + i Q(r)$ with real $r$; (ii) calculate the Sylvester matrix $S(P,Q)$ associated to $P(r)$ and $Q(r)$; (iii) if at least one of the principal minors of $S(P,Q)$ is negative or zero then $f(z)$ is unstable. 

Applying this procedure to the polynomial in (\ref{dispB}) we find the stability criteria 
\bea 
A_0 &>& 0 \\
A_2 &>& 0 ~~ \textrm{and} \\
A_2(A_1 A_2-A_0) &>&  \tilde{A}_1^2 ~~ .
\tea

Comparing this result to the isotropic case studied in \citet{balbus01}, we see that the stability criteria get modified by pressure anisotropy in two ways. Not only do the expressions for the coefficients $A_n$ contain $\alpha$, but also an extra term $\tilde{A}_1$ appears which vanishes if $\alpha = 1$ or if $b_x k_z = b_z k_x$. The first effect does not change in a significant way the quantitative analysis of \citet{balbus01}, so we shall not pursue it here.  
However, the appearance of $\tilde{A}_1$ in the third stability criteria implies that when pressure anisotropy is taken into account $A_1 A_2-A_0$ cannot be arbitrarily small, otherwise an instability develops. 
This represents a qualitative change with respect to the isotropic case, where $A_1 A_2-A_0 > 0$ for stable solutions.  Note that this modification only affects transverse perturbations, since $\tilde{A}_1 = 0$ if $k_\perp = 0$. 

\section{Application to the ICM}
\label{icm}
\begin{figure}[htb]
\begin{center}
\scalebox{1}{\includegraphics{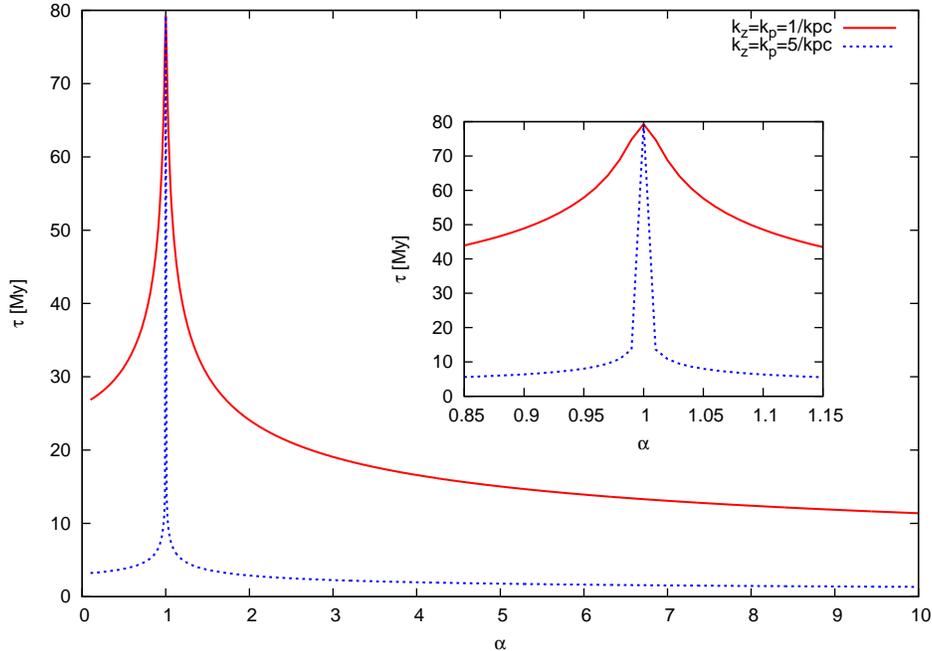}}
 \end{center}
\vspace{0.5cm}
\caption{(Color online) Growth time for the MTI/HBI as a function of $\alpha$ for $k_z=k_\perp=1/$kpc and $k_z=k_\perp=5/$kpc. The inset is a zoom for $0.85<\alpha<1.15$, which is a closer range to the values expected in the ICM.}
\label{bous1}
\end{figure}

As an illustrative case, we will now present numerical results corresponding to the ICM. As typical values we take $\rho \sim 10^{-3}$ cm$^{-3}$, $B \sim 10^{-6}$ G, $T \sim 10^{7}$ K, and $P \sim 10^{-10}$ ergs cm$^{-3}$. The value of $g$ is estimated as the average of $g(r)=G M/r^2$ from $r=r_c$ to $r=2r_c$, where $G$ is the gravitational constant, $M = 10^{14} M_\odot$ is the mass of the ICM and $r_c = 290$ kpc. This gives $g \sim 7 \times 10^{-7}$ cm s$^{-2}$. For the entropy gradient we take $3 d\ln s/dz = -d\ln T/dz$, and taking the Brunt--V\"ais\"al\"a frequency to be $\sim 10^{-12}$ s$^{-1}$ \citep{ruszkowski10} we get $d\ln T/dz \sim 10^{-23}$ cm$^{-1}$. 
The anisotropy parameter in the ICM can be written as $\alpha = 1/(1\pm \varphi)$, where $\varphi = |p_\perp-p_\parallel|/p_\perp\sim 1/\sqrt{\textrm{Re}}$ and Re is the Reynolds number \citep{Schekochihin05,rosin10,howes06}. To estimate the lower and upper bounds for the values of $\alpha$ we will consider Re $>$ 50, so we get $\alpha \in [0.87,1.16]$. Note that in the ICM Re $\gg$ 50 for convective motions, so that actually even lower values for $|1-\alpha|$ are expected in this environment. However, in order to show the physical behaviour of this instability as a function of the anisotropy parameter it is convenient to consider a broader, though unphysical, range for $\alpha$ (as we do in Figure \ref{bous1}). In what follows we take a tangled magnetic field corresponding to $b_x=0.3$.  

In order to illustrate the dependence of the stability of solutions to (\ref{dispB}) on pressure anisotropy, we show in Figure \ref{bous1} the growth time $\tau=1/|\textrm{Im}(\omega)|$ as a function of $\alpha$ for $k_z=k_\perp=1/$kpc and $k_z=k_\perp=5/$kpc. 

It is seen from Figure \ref{bous1} that in the case of $\alpha=1$ the growth time, $\tau$, is of the order of $\sim 100$ My \footnote{The scale dependence of $\tau$ arises from considering a finite conductivity. However, this scale dependence does not affect the order of magnitude of $\tau$.}. On the othe hand, the presence of anisotropy causes a strong decrease in $\tau$, which is considerably larger for smaller scales. The dependence of $\tau$ with $\alpha$ is similar in both cases ($\alpha<1$ or $\alpha>1$). The inset shows the behaviour of $\tau$ in the range $0.85<\alpha<1.15$,  which as mentioned is a closer range to the values expected in the ICM. It is seen that for scales $\sim 0.2$ kpc the growth time in the anisotropic case can be almost one order of magnitude smaller than the one corresponding to $\alpha=1$. For larger scales ($>1$ kpc) $\tau$ decreases by a factor of $\sim 2$ or less.      

\begin{figure}[htb]
\begin{center}
\scalebox{1}{\includegraphics{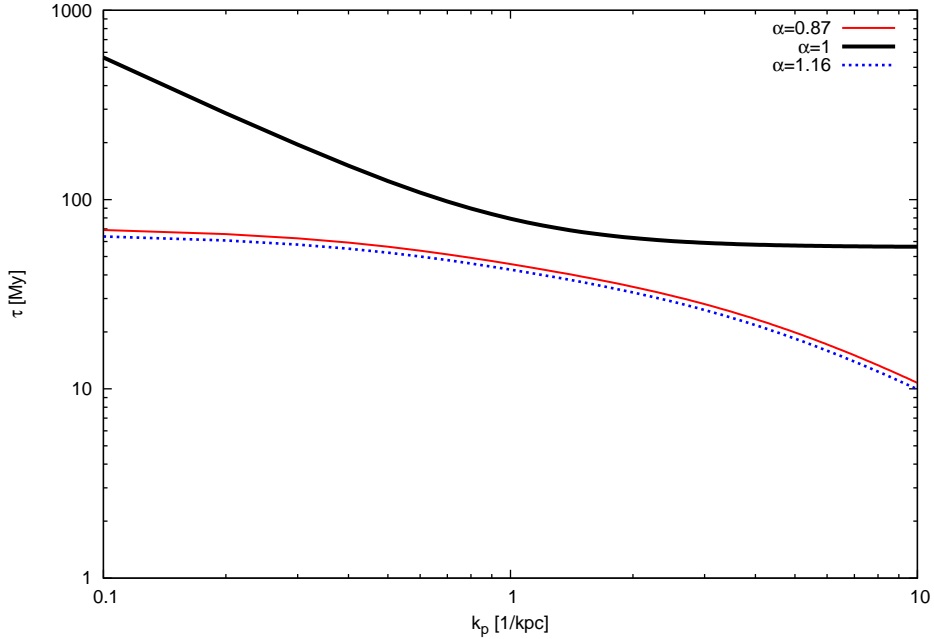}}
 \end{center}
\vspace{0.5cm}
\caption{(Color online) Growth time for the MTI/HBI as a function of $k_\perp$ for $k_z=1/$kpc and $\alpha=0.87;1;1.16$.}
\label{bous1kz}
\end{figure}

We now go over to analise the dependence of $\tau$ with wave vector in the range $[0.1,10]\times 1/$kpc (magnetic tension stabilizes the instabilities on shorter scales). Figure \ref{bous1kz} shows $\tau$ as a function of $k_\perp$ for $k_z = 1/$kpc and for three values of $\alpha$. We note that for other values of $k_z$ in the studied range $[0.1,10]\times 1/$kpc the results are similar. It is seen that for any value of $k_\perp$ the growth time of the isotropic case is larger than the corresponding ones to $\alpha<1$ and $\alpha>1$. The difference in $\tau$ can reach one order of magnitude. As happened in the previous case, the behaviour of $\tau$ is similar for both anisotropic cases. Interestingly, the decrease in $\tau$ with the anisotropy is significantly smaller in the kpc range. 

We will now briefly discuss some possible implications of our findings to the ICM (in the case where the timescale ordering of eq. (\ref{ord}) does not apply). In this connection, the most important results of this work are that: (i) even with small pressure anisotropy the growth time of the MTI/HBI can become almost an order of magnitude smaller than the one corresponding to the isotropic case; (ii) the decrease of $\tau$ with anisotropy is smaller for scales $\sim$ kpc; and (iii) the effect of anisotropy on $\tau$ is larger at smaller scales. As mentioned in the Introduction, the MTI/HBI can be a significant source of magnetic field amplification \citep{bu10,mccourt10,parrish07}. Our results imply that in the ICM the magnetic field amplification due to the MTI/HBI could be faster than what is expected in the isotropic case. Although we expect this effect to be small in the ICM because the magnetic field is weak and thus the timescale ordering holds, it could have some important consequences in the subsequent plasma dynamics following the linear regime.

\section{Conclusions}
\label{con}

In this work we have studied the influence of pressure anisotropy, which is present in dilute magnetized space plasmas, on the physics of the MTI and the HBI. We have performed a linear analysis based on fluid equations which go beyond the CGL double--adiabatic formalism. 

The main conclusion that we can extract from our results is that, for the conditions prevailing in the ICM, the impact of pressure anisotropy on the MTI and HBI is small. More specifically, if the magnetic field is sufficiently weak so that the dynamical frequency is larger than any other frequency of the system, these instabilities do not depend on pressure anisotropy (provided the latter is small). 

On the other hand, if this timescale ordering does not apply the stability criteria for the MTI and the HBI will depend on pressure anisotropy in two ways, first through a dependence of the terms which also appear in the isotropic case, and second through the appearance of new terms. We find that these extra terms affect the stability of the stratified plasma and the growth rate of the MTI/HBI. Specifically, we find that the growth time of the instability in the anisotropic case can be almost one order of magnitude smaller than the isotropic one. 

The analysis of the MTI/HBI in anisotropic plasmas presented in this work is largely idealized, since we limited ourselves to the linear regime, we neglected viscosity and radiative cooling \citep{balbus10}, we considered static pressure anisotropy configurations (on which the MTI/HBI develops), and finally we used a simplified model of the ICM. In spite of this shortcomings, we believe that our analysis provides some insight, at least preliminary, into the effect of pressure anisotropy on the MTI/HBI, as well as on some possible implications for the dynamics of the ICM and similar astrophysical environments.

\section*{Note added in v3}

Shortly after this work was submitted to the ArXiv, a preprint by M. Kunz \citep{KUNZ} appeared analyzing the impact of pressure anisotropy on the MTI/HBI by using the Braginskii anisotropic viscosity equation for $p_\perp-p_\parallel$. In \citep{KUNZ}, the author introduces time--scale and amplitude orderings measured by the Mach number $M\ll 1$, whereby the Boussineq approximation is implemented as $\delta \rho/\rho \sim \delta T/T \sim \delta p/(Mp)$ (i.e. relative changes in the pressure are much smaller than relative changes in the temperature or density), and not as $\delta p_\perp = \delta p_\parallel \sim 0$ as we naively assumed here. 

This naive interpretation of the Boussineq approximation eliminates the Braginskii viscosity, and therefore in our work pressure anisotropy is not self--consistently calculated but rather acts as a (fixed) background on which the MTI/HBI develop. Due to this, our results differ from those of \citep{KUNZ} in that we find, for the conditions prevailing in the ICM, only moderate changes in the MTI/HBI growth rates when including pressure anisotropy instead of the drastic changes  found and described in \citep{KUNZ}. This clearly highlights the necessity of including Braginskii viscosity {\it self--consistently} in order to study the effect of pressure anisotropy on the MTI/HBI as taking place in the ICM.

In view of the above, we believe that the approximations involved in this work should be regarded merely as a first step towards a more thorough understanding of the MTI and HBI within the 16--moment formalism for collisionless magnetized plasmas. Concerning this, the relation between the equation for the pressure anisotropy used in \citep{KUNZ} and the evolution equations for $p_\perp$ and $p_\parallel$ of the 16--moment formalism is not clearly established. Hence it is still not clear what may be the consequences of the time--scale and amplitude orderings mentioned above on the 16--moment equations. We are currently investigating these issues.

\begin{acknowledgements}
      This work was partially funded by Funda\c c\~ao de Amparo a Pesquisa do Estado de S\~ao Paulo (FAPESP -- Brazil). We are very grateful to Mike McCourt and to Matthew Kunz for many illuminating comments and discussions, and to Elisabete de Gouveia Dal Pino for her useful suggestions.  
\end{acknowledgements}

\bibliographystyle{aa}
\bibliography{mp} 
 {\typeout{}
  \typeout{****************************************************}
  \typeout{****************************************************}
  \typeout{** Please run "bibtex \jobname" to optain}
  \typeout{** the bibliography and then re-run LaTeX}
  \typeout{** twice to fix the references!}
  \typeout{****************************************************}
  \typeout{****************************************************}
  \typeout{}
}

\end{document}